\documentclass{elsart}
\usepackage{epsfig}
\usepackage{amssymb}
\begin{document}
\begin{frontmatter}

\title{Novel order parameter to describe
the critical behavior of Ising spin glass models}

\author{F. Rom\'a, F. Nieto, A. J. Ramirez-Pastor\corauthref{cor1}}
\address{Departamento de F\'{\i}sica, Universidad Nacional de San Luis - CONICET. Chacabuco 917, 5700 San Luis,
Argentina. \\
 E-mail: froma@unsl.edu.ar, fnieto@unsl.edu.ar, antorami@unsl.edu.ar}

\author{E.E. Vogel}

\address{Departamento de F\'{\i}sica, Universidad de La
Frontera, Casilla 54-D, Temuco, Chile. \\ E-mail:
ee{\_}vogel@ufro.cl}

\thanks[cor1]{Corresponding author. Fax +54-2652-430224, E-mail: antorami@unsl.edu.ar}

\begin{abstract}

A novel order parameter $\Phi$ for spin glasses is defined based
on topological criteria and with a clear physical interpretation.
$\Phi$ is first investigated for well known magnetic systems and
then applied to the Edwards-Anderson $\pm J$ model on a square
lattice, comparing its properties with the usual $q$ order
parameter. Finite size scaling procedures are performed. Results
and analyses based on $\Phi$ confirm a zero temperature phase
transition and allow to identify the low temperature phase. The
advantages of $\Phi$ are brought out and its physical meaning is
established.

\end{abstract}

\end{frontmatter}
\parskip 0cm

\section{Introduction}

Despite over three decades of intensive work, the nature of the
low temperature phase of two-dimensional Edwards-Anderson (EA)
\cite{EA} model for spin glasses remains controversial. It is
agreed that a phase transition occurs at zero temperature  for a
Gaussian distribution of bonds (GD)
\cite{McMillan,Bray,Rieger,Hartmann}. Similarly, for a symmetric
$\pm J$ distribution or bimodal distribution (BD) of bonds, very
convincing numerical evidence has been found that there is no
transition at finite temperature
\cite{Hartmann,Bhatt,Kawashima,Houdayer,Hartmann2,Amoruso,Lukic}.
In most of these references, the authors do not use an order
parameter for characterizing the phase transition. On the other
hand, data arising from other contributions, which are based on
the behavior of a standard overlapping order parameter, support
the existence of a finite critical temperature
\cite{Shirakura,Shiomi,Matsubara}.

In this context, the main purposes of this paper are the
following: a) To show that the disagreement pointed out in
previous paragraph is related to the non-zero overlap of
site-order parameters obtained for quite distinct energy valleys;
b) To overcome this situation by proposing here a novel order
parameter $\Phi$, which is quite drastic to characterize phases
but still is general enough to coincide with usual descriptions of
ferromagnetic (F) and antiferromagnetic (AF) systems; c) To apply
$\Phi$ to do a scaling analysis for two-dimensional EA systems
including Binder cumulant \cite{Binder01}; d) To confirm the
assumption of the zero-temperature phase transition for
two-dimensional BD, thus reinforcing this result obtained by
previously quoted authors; and e) To give a physical meaning to
this result by using the grounds on which $\Phi$ is based on.

The present work is organized as it follows. In Section 2, we
introduce the model and define a novel order parameter, $\Phi$,
very useful for spin glasses and other frustrated systems. Results
of the simulation are presented in Section 3. Finally, our
conclusions are drawn in Section 4.

\section{Model and basic definitions}

Let us begin by very briefly introducing the system under study.
Ising spin $s_i$ occupies $i-th$ site of a two dimensional (square
for simplicity) lattice. The interaction with the spin at site $j$
is mediated by exchange interaction $J_{ij}$. In the absence of
magnetic field (which is the case for the scope of the present
paper) the Hamiltonian of such system can then be written as
\begin{equation}
H = \sum_{\langle i,j \rangle} J_{ij} s_is_j \; ,
\end{equation}
where interactions $\{J_{ij}\}$ are restricted to nearest neighbor
couplings. In the ferromagnetic (F) Ising model, $J_{ij}=-J$
$\forall$ $\langle i,j \rangle$. For the EA model, we will
consider half of the bonds F, while the other half will be
described by antiferromagnetic (AF) bonds of the same magnitude,
namely, $J_{ij}=+J$ ($J>0$). A sample is one of the possible
random distributions of these mixed bonds. For simplicity spins
take values $s_j=\pm 1$, which can be equally denoted by their
signs.


Now, let us consider a configuration $\alpha$ defined by a
collection of ordered spin orientations $\{s^{\alpha}_j\}$. The
usual EA order parameter $q$ is built up by means of overlaps
between two configurations $\alpha$ and $\beta$ and takes the form
\begin{equation}
q_{\alpha\beta}=\frac{1}{N}\sum _{j=1} ^{N} s_j ^{\alpha}s_j
^{\beta}, \label{eqqab}
\end{equation}
where $N$ ($\equiv L \times L$) is the total number of spins.

For models in which the ground state is non degenerate after
breaking ergodicity, such as the pure F case, the distribution of
$q_{\alpha\beta}$ values for the ground manifold (T=0.0) is
trivial and it is given by delta functions at $q_{\alpha\beta} =
1.0$  and $q_{\alpha\beta}=-1.0$. This also happens in general for
all systems with non-degenerate ground level. But this also
applies to GD, where local fields have all different values at
different sites, leading to a true minimum energy for just one
pair of opposite ground states. However, for the BD the local
field assumes a few discrete values only, which necessarily means
highly degenerate ground manifolds leading to $|q_{\alpha\beta}|
<1.0$, for a large number of possible pairs of ground states. This
distribution will have two broad symmetric maxima but it will not
vanish in the intermediate region \cite{Shiomi}.

On the other hand, a more detailed description based on a
topological picture of the ground state of BD was presented
\cite{Vogel01,Vogel02}. This framework allows us to define a state
function with a clear physical meaning, which is a good candidate
to be a new order parameter for a phase transition. In fact, it
has been reported an important feature of the ground state,
namely, at $T=0$ there exist clusters of solidary spins (CSS)
preserving the magnetic memory of the system (solidary spins
maintain their relative orientation for all states of the ground
manifold)\cite{Barahona1982}. The main idea of this work is to
characterize the nature of the low temperature phase through the
CSS.

Let us consider a particular sample of any given size $N$. We
denote by $\Gamma_{\kappa}$ any of the $n$ CSS of the sample
($\kappa$ runs from $1$ to $n$). Calculations begin recognizing
all of the CSS of each sample belonging to a set of $2000$
randomly generated samples of each size. This process is closely
related to finding the so-called ``diluted lattice" that prevails
after removing all frustrated bonds \cite{Valdes98}, so the
algorithms designed for that purpose can also be used here.

Let us first pick any arbitrary ground state configuration denoted
by an asterisk ($*$) fixing one of the two possible relative
orientations of the CSS, thus becoming a reference configuration.
Then a local overlap corresponding to the configuration $\alpha$
in the ${\kappa}$-th cluster, of size $N_{\kappa}$, can be defined
as

\begin{equation}
\phi _{\kappa}^{\alpha} =\frac{1}{N_{\kappa}} \sum _{j \in \Gamma
_{\kappa}} s^{*}_{j} s_j^{\alpha}, \label{lop}
\end{equation}
where the sum runs over all spins in the cluster $\Gamma_{\kappa}$
only. Thus,  $\vert \phi_{\kappa}^{\alpha} \vert =1$ indicates a
fully ordered cluster; otherwise $\vert \phi_{\kappa}^{\alpha}
\vert < 1$. The magnetic order of the sample is characterized by
the set of overlaps, namely, $\{ \phi_{\kappa}^{\alpha} \}$. Under
the occurrence of a phase transition, the new set $\{
\phi_{\kappa}^{\alpha} \}$ will determine uniquely the ergodic
component of the reached phase. This fact is a required
characteristic for a well behaved order parameter \cite{Enter84}.

We are now ready to define the new order parameter introduced in
this paper. It is given by
\begin{equation}
\Phi_{\alpha} =  \sum _{{\kappa}=1}^n f_{\kappa} \vert
\phi_{\kappa}^{\alpha} \vert, \label{op}
\end{equation}
where $f_{\kappa}=N_{\kappa}/N_I$, being $N_I=\sum_{{\kappa}=1}^n
N_{\kappa}$, ($N_I\le N$). From the definition it flows that for
$T \geq T_c$ the average value of $\Phi_{\alpha}$, namely, $\Phi$,
should be $0$. Similarly, for $T < T_c$ it should hold that
$0<\Phi \leq 1$, being $\Phi = 1$ for $T=0$ only. It is important
to emphasize that $\Phi_{\alpha}$ is a state function, which is an
advantage over $q_{\alpha\beta}$ defined in eq. (\ref{eqqab}) as
an overlap between two configurations of the system.

The calculation of the new order parameter $\Phi_{\alpha}$
requires the previous determination of the set of CSS for each
considered sample. This procedure, which was performed by using
the numerical scheme introduced in Ref.\cite{Ramirez04}, is a
computational limitation for going to larger system sizes. Once
the ground manifold of each sample is completely characterized
after this procedure, the numerical calculations converge very
quickly by flipping the spins not present in the largest CSS only.
The second run on each sample takes much less time than the first
one that is needed to find all CSS.

In the F Ising model there is a unique cluster of $N$ solidary
parallel spins at $T=0$. As it can be trivially demonstrated, eq.
(\ref{op}) leads to the magnetization per spin, which is the
natural order parameter of such system. Similarly, for the AF case
we get the well-known order parameter defined as the magnetization
difference between the two possible interpenetrating sublattices.
Finally, for GD, there also exists an unique CSS in the ground
state and the phase transition occurring in the system is
completely described by the new order parameter, eq.(\ref{op}), as
well. So the new parameter retains all the properties of the
well-known non-degenerate systems.

Finally, the reduced fourth-order cumulant, introduced by Binder
\cite{Binder01}, can be calculated as
\begin{equation}
U_L=1-\frac{\left[ \langle m^4 \rangle \right]}{3\left[ \langle
m^2 \rangle \right] ^2}, \label{Bcumulant}
\end{equation}
where $m$ is a given order parameter, and $\langle \ldots \rangle$
and $[ \ldots ]$ mean the spin configuration (thermal) average and
the bond configuration (sample) average, respectively. In general,
the structure of the distribution of $m$ affects the behavior of
the fourth-order Binder cumulant. Thus, for a trivial
distribution, both $|m| \to \pm 1$ and $U_L \to 2/3$, as $T$ goes
to zero. On the other hand, if the distribution is nontrivial,
$|m|$ tends to a value $m^o$ lower than 1, while $U_L$ tends to a
value $U_L^o$ lower than $2/3$ upon decreasing temperature.

\section{Results}

Distribution functions for  $q_{\alpha \beta}$ and
$\Phi_{\alpha}$, were obtained for BD by using a standard
simulated-tempering procedure\footnote{It must be emphasized that
it is not necessary a simulated-tempering scheme for calculating
the new order parameter, $\Phi$.} \cite{marinari,kerler} along
with the well known Glauber's dynamics \cite{kawa72}. For
illustration purposes, we perform calculations on $1000$ samples
of size 64 $(8 \times 8)$ at different temperatures ranging from
$T=0.2$ to $T=1.0$. (throughout this paper, $k_B/J=1$ without any
loss of generality). The results corresponding to $T=0.31$,
$T=0.53$ and $T=0.69$ are presented in Fig. 1. As it is shown in
part (a), the distribution of the new order parameter,
$R(\Phi_{\alpha})$, exhibits a drastic behavior as $T$ decreases.
In part (b) it is shown how the corresponding curves for
$r(|q_{\alpha \beta}|)$ have a broad maximum over the plotted
range and $r(0)>0$. These undesired characteristics for this order
parameter remain even at low temperatures.

In Fig. 2, $U_L(T)$, built up from $R(\Phi_{\alpha})$
distribution, is presented for different lattice sizes ranging
from $N=16$ to $N=144$ and each point was calculated by averaging
over a set of 2000 samples. With the help of the inset, it is
observed that the curves do not intersect each other as a direct
indication of the absence of a phase transition for finite
temperature, at least for the sizes considered here. Eventually we
are not free from finite size considerations yet as it has been
recently proposed that at least samples with $L=50$ should be
reached when conventional parameters are used \cite{Katzgraber05}.
However, using a more drastic parameter like the one proposed
here, a faster convergence towards large $L$ values is expected.
It is clear that all curves go to $2/3$ as $T \to 0$, which
reinforces the robustness of eq.(\ref{op}). On the other hand,
this property is not followed by cumulants obtained from other
overlapping order parameters. This is the case of Fig. 7 in
Ref.\cite{Shiomi}, where it is possible to think that the reported
crossing of the cumulants of $q$ arises from the dependence of
$U_L^o$ on size. In this contribution, the authors reported a
critical temperature different from zero, $T_c \approx 0.23$.

Finite-size scaling \cite{Binder01} predicts that all curves in a
figure such as Fig. 2, should collapse onto a single one when
using $(T-T_c)L^{1/\nu}$ as independent variable, being $T_c$ the
critical temperature for the transition and $\nu$ an appropriate
critical exponent. Upon choosing $T_c=0$ and the exponent $\nu$ is
taken as $\nu=2.63 \pm 0.20$, the standard universal behavior for
$U_L(T)$ is obtained as shown in Fig. 3. This is an independent
confirmation of previously reported results
\cite{Bhatt,Young01,Cheung,McMillan01}.

 The following two parameters were also measured as each
sample was solved exactly: (a) The mean fraction of spins, $P
\equiv \left[ N_I \right] /N$, belonging to any CSS; and (b) The
fraction of spins in the largest CSS $p\equiv \left[ N_{\ell}
\right] /N$, where $N_{\ell}$ is the number of spins in the
largest CSS.  Fig. 4(a) shows that while $P$ remains rather
constant, $p$ clearly decreases with size and the stabilizing role
of the largest CSS is lost. The average number of CSS $[n]$ as
function of size was also measured, finding that $[n]$ grows
linearly with $N$, as it is shown in Fig. 4(b). For $N>49$, say
(when small size effects do not play an important role), the
following approximate law is obtained $[n]\approx 0.03N+0.60$.

The size dependence of a possible spin-glass phase can be
described in the following terms. For small sizes ($N<49$ say)
most of solidary spins are grouped in one large cluster
stabilizing a spin-glass phase. As size grows, the number of CSS
increases linearly with $N$ (or quadratically with $L$), while the
relative size of the largest CSS diminishes. This can be
visualized as if the original lattice would break into portions of
relatively smaller sizes, none of them large enough to stabilize a
spin-glass phase. This is the reason for the numeric result of
Fig. 2, showing no intersection of curves for different sizes.

If the same procedure used here for the symmetric case is applied
to different concentrations of F and AF bonds, a stable phase is
found in the extremes of high and low concentrations of F bonds in
correspondence with results already reported in the literature
\cite{Mor01,Mor02}. As the relative concentration of F bonds
varies the behavior of $P$ and $[n]$ is very similar to that shown
in Fig.~4. However, $p$ tends to be constant for very asymmetric
distributions of $\pm J$ bonds. The last statement indicates the
presence of an infinite CSS in the thermodynamical limit, which is
associated to a stable phase. These results are not shown
graphically in the present paper.

\section{Conclusions}

A new order parameter $\Phi$ has been introduced and applied to
the study of magnetic systems. It proves to be particularly
essential for characterizing degenerate systems such as Ising-like
models with bimodal distribution. Parameter $\Phi$ is well
behaved, having all desired properties for a drastic order
parameter. This behavior is based on the properties of CSS. When
this order parameter is used for systems with BD, properties
similar to order parameters for non-degenerate systems are found.
Then, the characterization of magnetic phases after using the
scaling techniques of cumulants becomes unambiguous. In this way
it was shown that the two-dimensional Edwards-Anderson model
exhibits a phase transition at $T_c=0$, with a critical exponent
$\nu=2.63 \pm 0.20$.

The identification of all CSS for each sample is the bottleneck in
the present computational scheme. This procedure is very time
consuming for large lattice sizes. The extra time needed for
finding all CSS is well paid by the better precision achieved in
the characterization of the phase, and the elimination of overlaps
in the new order parameter, thus making the identification of the
ergodic valley reliable.

The characteristics of this new order parameter make it also
useful for other frustrated systems, where large overlaps occur
due to the complex energy valley. The extension of the use of the
parameter $\Phi$ to other kind of problems is clearly foreseen.
For instance, it can be the key element $i)$ to describe the phase
diagram for the asymmetric distribution problem around the
critical concentration of ferromagnetic (or antiferromagnetic)
bonds of $\approx 0.1$ and $ii)$ to study the critical behavior of
3D Ising spin glasses. This task requires serious improvements in
the numerical techniques used in order to get access to large
lattice sizes. Work along this line is in progress.

\ \

\noindent{\bf Acknowledgments}

\ \

We thank Fondecyt (Chile) under projects 1020993 and 7020993. One
author (EEV) thanks Millennium Scientific Initiative (Chile) under
contract P-02-054-F for partial support. Three authors (FR, FN and
AJRP) thank CONICET (Argentina) and the Universidad Nacional de
San Luis (Argentina) under project 322000.

\newpage

\begin{center}
\section*{FIGURE CAPTIONS}
\end{center}

\vspace{1cm}

\noindent Fig. 1. Distributions of the order parameters (a)
$\Phi_{\alpha}$ and (b) $|q_{\alpha \beta}|$,
at 3 different temperatures as indicated. \\[18pt]

\noindent Fig. 2. Cumulant $U_L(T)$ plotted versus  $T$ for
various lattice sizes as indicated. The inset zooms
the area indicated by a dashed frame.\\[18pt]

\noindent Fig. 3. Scaling plot of $U_L$ against
$(T-T_c)L^{1/\nu}$, with $T_c=0$ and $\nu=2.63 \pm 0.20$.
\\[18pt]

\noindent Fig. 4. (a) $P$ and $p$ and (b) the growing of $[n]$ as
a function of the lattice size, respectively.

\newpage
\begin{figure}[tbp]
\centering
\includegraphics [width=20cm,angle=0]{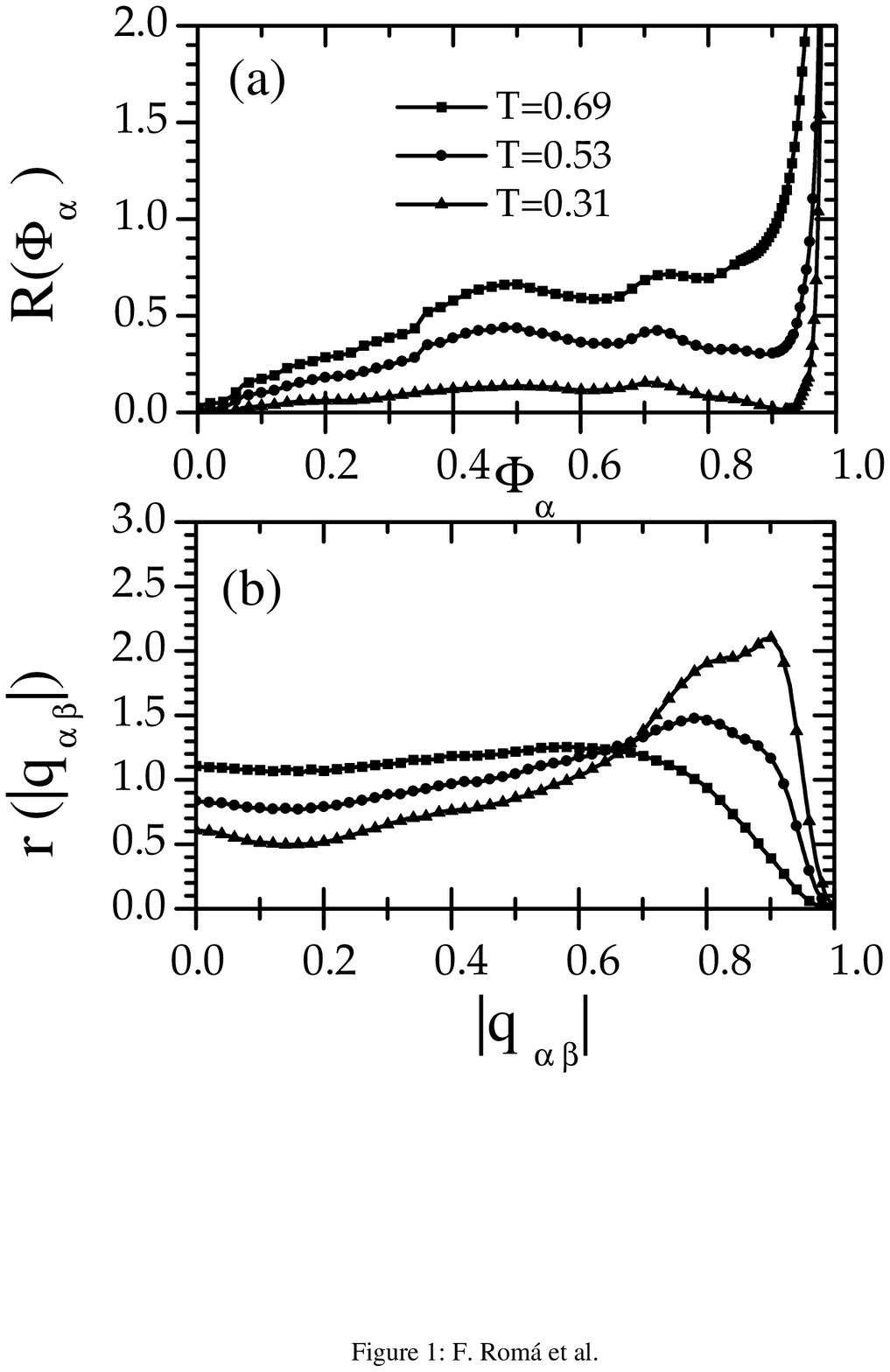}
\end{figure}

\newpage
\begin{figure}[tbp]
\centering
\includegraphics [width=30cm,angle=0]{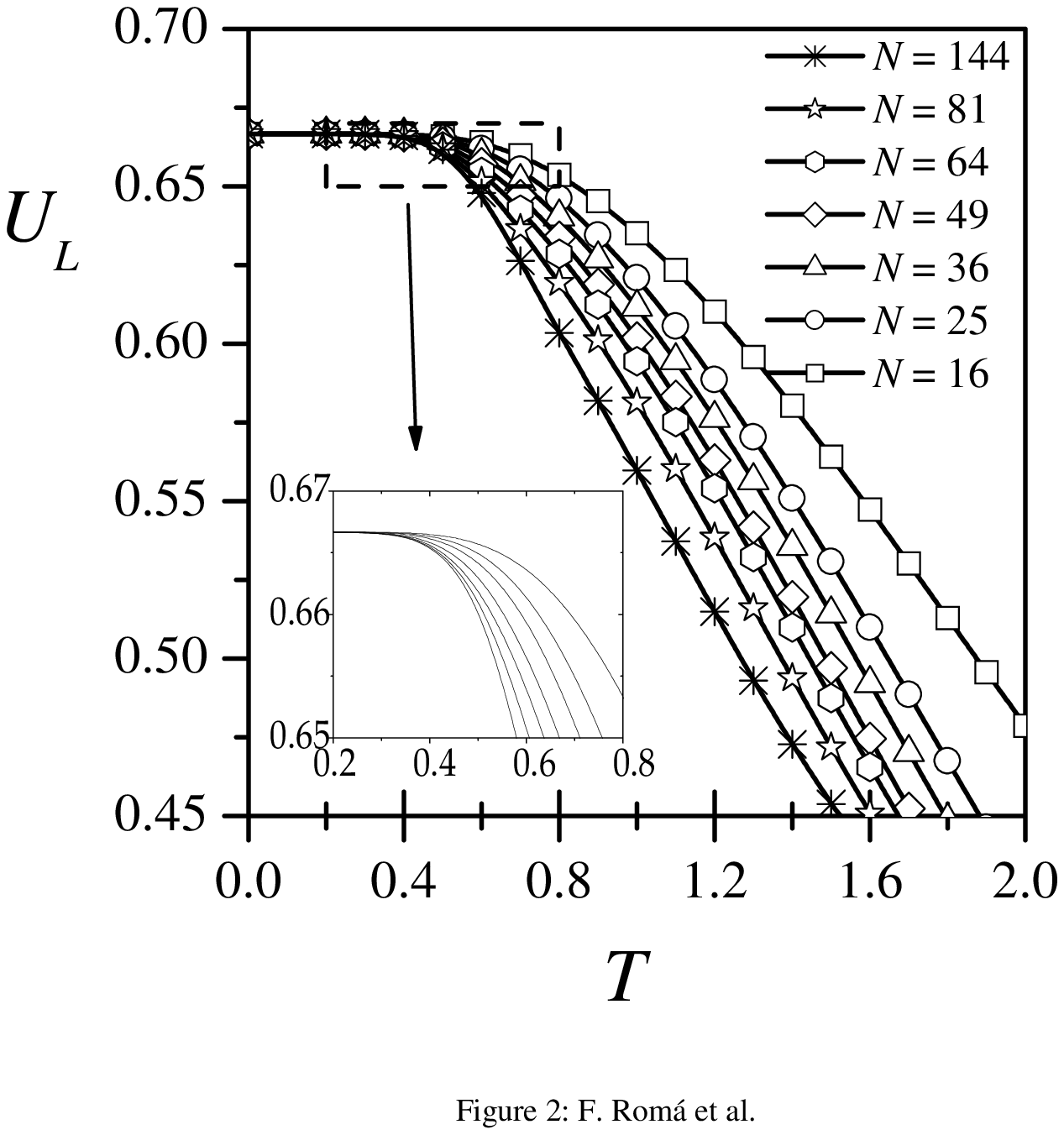}
\end{figure}

\newpage
\begin{figure}[tbp]
\centering
\includegraphics [width=30cm,angle=0]{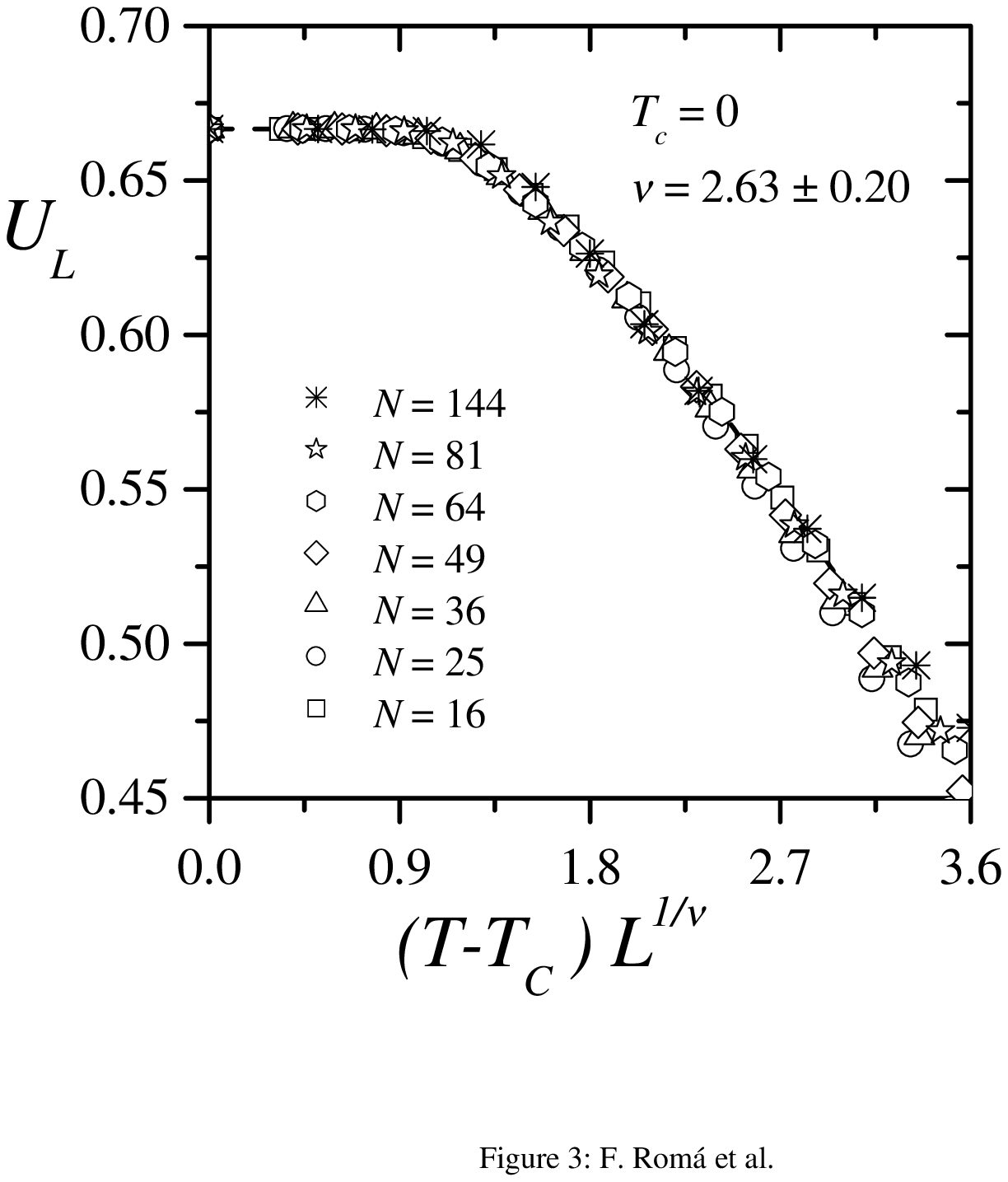}
\end{figure}

\newpage
\begin{figure}[tbp]
\centering
\includegraphics [width=20cm,angle=0]{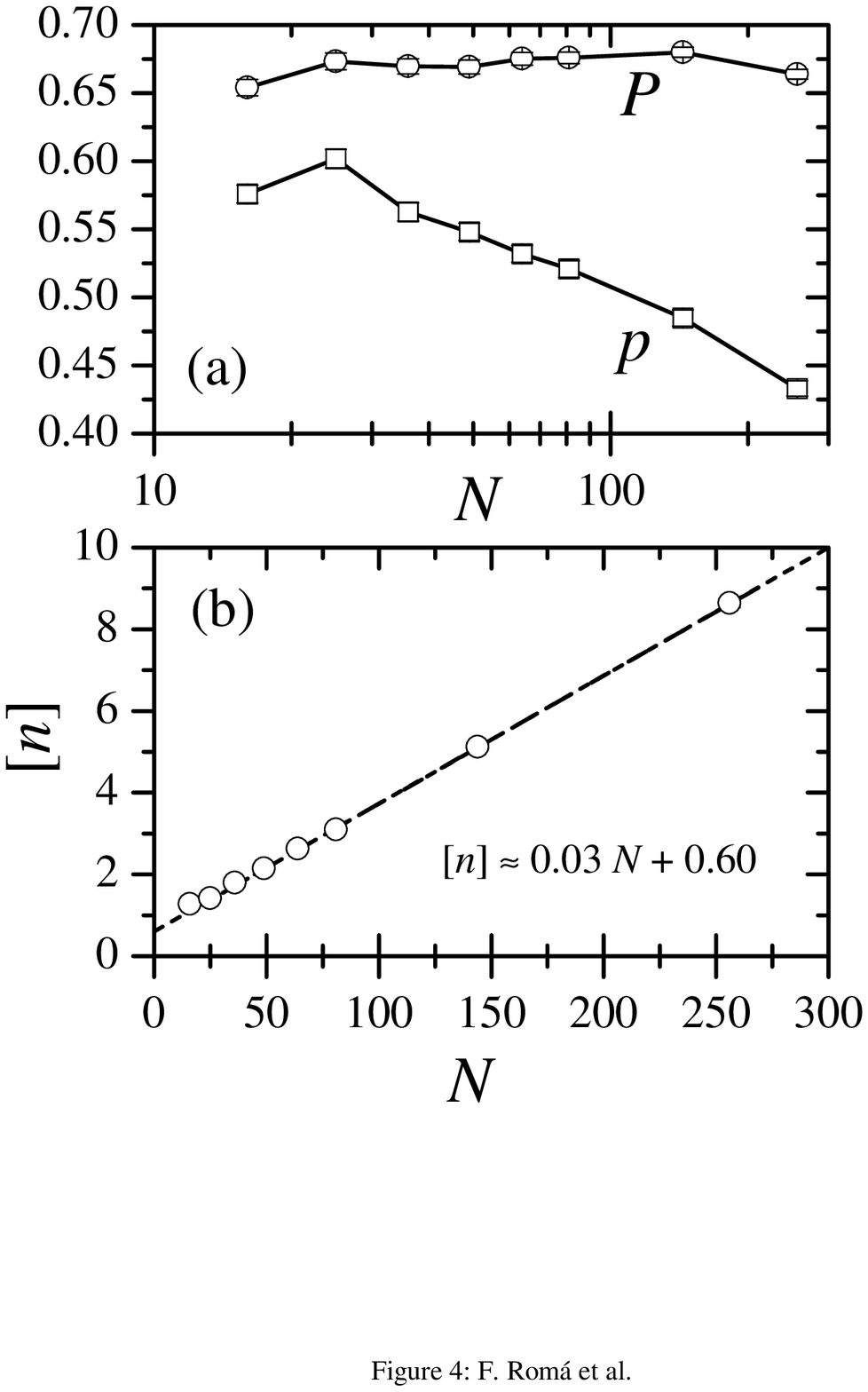}
\end{figure}



\newpage

\begin{thebibliography}{9}

\bibitem{EA} S.F. Edwards and P.W. Anderson, J. Phys. F \bf 5, \rm 965 (1975); G. Toulouse, Commun.
Phys. \bf 2, \rm 115 (1977).
\bibitem{McMillan} W.L. McMillan, Phys. Rev. B \bf 30, \rm R476 (1984).
\bibitem{Bray} A.J. Bray and M.A. Moore, J. Phys. C \bf 17, \rm L463 (1984); A.J. Bray
and M.A. Moore, in \em Heidelberg Colloquium on Glassy Dynamics,
\rm edited by J.L. van Hemmen and I. Morgenstern (Springer-Verlag,
Heidelberg, 1987).
\bibitem{Rieger} H. Rieger, L. Santen, U. Blasum-U, M. Diehl, M. J\"unger, and G.
Rinaldi, J. Phys. A \bf 29, \rm 3939 (1996).
\bibitem{Hartmann} A. K. Hartmann and A. P. Young, Phys. Rev. B \bf 64, \rm 180404(R) (2001).
\bibitem{Bhatt} R. N. Bhatt and A. P. Young, Phys. Rev. B \bf 37, \rm 5606 (1988).
\bibitem{Kawashima} N. Kawashima and H. Rieger, Europhys. Lett. \bf 39, \rm 85 (1997).
\bibitem{Houdayer} J. Houdayer, Eur. Phys. J. B \bf 22, \rm 479 (2001).
\bibitem{Hartmann2} A. K. Hartmann, A. J. Bray, A. C. Carter, M. A. Moore, A. P. Young, Phys. Rev. B \bf 66 \rm 224401 (2002).
\bibitem{Amoruso} C. Amoruso,  E. Marinari, O. C. Martin, A. Pagnani, Phys. Rev. Lett. \bf 91, \rm 087201 (2003).
\bibitem{Lukic} J. Lukic, A. Galluccio, E. Marinari, O. C. Martin, G. Rinaldi, Phys. Rev. Lett. \bf 92, \rm 117202 (2004).
\bibitem{Shirakura}  T. Shirakura and F. Matsubara, Phys. Rev. Lett. \bf 79, \rm 2887 (1997).
\bibitem{Shiomi} M. Shiomi, F. Matsubara and T. Shirakura, J. Phys. Soc. Jpn. \bf 69, \rm 2798 (2000).
\bibitem{Matsubara} F. Matsubara, T. Shirakura and M. Shiomi, Phys. Rev. B \bf 58, \rm R11821 (1998).
\bibitem{Binder01} K. Binder, Rep. Prog. Phys. \bf 60, \rm 488 (1997).
\bibitem{Vogel01} E.E. Vogel, S. Contreras, M.A. Osorio, J. Cartes,
F. Nieto and A.J. Ramirez-Pastor, Phys. Rev. B \bf 58, \rm 8475 (1998).
\bibitem{Vogel02} E.E. Vogel, S. Contreras, F. Nieto and A.J. Ramirez-Pastor,
Physica A \bf 257, \rm 256 (1998).
\bibitem{Barahona1982} F. Barahona, R. Maynard, R. Rammal and J.P.
Uhry, J. Phys. A: Math. Gen. \bf 15, \rm 673 (1982).
\bibitem{Valdes98} J.F. Valdes, J. Cartes, E.E. Vogel, S. Kobe,
and T. Klotz, Physica A \bf 257, \rm 557 (1998).
\bibitem{Enter84} A.C.D. van Enter and J.L. van Hemmen, Phys. Rev. A {\bf 29}, 355 (1984).
\bibitem{Ramirez04} A.J. Ramirez-Pastor, F. Rom\'a, F. Nieto and E.E. Vogel, Phys. A {\bf 336}, 454 (2004).
\bibitem{marinari} E. Marinari and G. Parisi, Eurphys. Lett. {\bf
19},
451 (1992).
\bibitem{kerler} W. Kerler and P. Rehberg, Phys. Rev. E {\bf 50}, 4220 (1994).
\bibitem{kawa72} K. Kawasaki, in {\it Phase Transitions and Critical Phenomena},
edited by C. Domb and M. Green  (Academic Press, London, 1972),
Vol. 2.
\bibitem{Katzgraber05} H.G. Katzgraber and L.W Lee, Phys. Rev. B \bf 71, \rm 134404 (2005).
\bibitem{Young01} A. P. Young, J. Phys. C \bf 17, \rm L517 (1984).
\bibitem{Cheung} H.F. Cheung and W.L. McMillan, J. Phys. C \bf 16, \rm 7027 (1983).
\bibitem{McMillan01} W.L. McMillan, Phys. Rev. B \bf 28, \rm 5216 (1983).
\bibitem{Mor01} I. Morgenstern and K. Binder, Phys. Rev. Lett. \bf 43, \rm 1615 (1979).
\bibitem{Mor02} I. Morgenstern and K. Binder, Phys. Rev. B \bf 22, \rm 288 (1980).
\end{thebibliography}
\end{document}